\documentclass[twocolumn,secnumarabic,amssymb,nobibnotes,aps, prd]{revtex4-1}

\usepackage{graphicx}
\usepackage{hyperref}
\hypersetup{colorlinks,citecolor=blue}
\usepackage{amsmath}
\usepackage{xcolor}

\begin{document}
\title{Color-flavor locked quark stars in energy-momentum squared gravity}%

\author{Ksh. Newton Singh}%
\email[Email:]{ntnphy@gmail.com}
\affiliation{Department of Physics, National Defence Academy, Khadakwasla, Pune-411023, India. \\
Department of Mathematics, Jadavpur University, Kolkata-700032, India}

\author{Ayan Banerjee}%
\email[Email:]{ayan\_7575@yahoo.co.in,}
\affiliation{Astrophysics and Cosmology Research Unit, University of KwaZulu Natal, Private Bag
X54001, Durban 4000, South Africa,}

\author{S. K. Maurya}%
\email[Email:]{sunil@unizwa.edu.om}
\affiliation{Department of Mathematical and Physical Sciences,
College of Arts and Sciences, University of Nizwa, Nizwa, Sultanate of Oman}

\author{Farook Rahaman}%
\email[Email:]{rahaman@iucaa.ernet.in}
\affiliation{Department of Mathematics, Jadavpur University, Kolkata-700032, India}

\author{Anirudh Pradhan}%
\email[Email:]{ pradhan.anirudh@gmail.com}
\affiliation{Department of Mathematics, Institute of Applied Sciences and Humanities, GLA University, Mathura-281 406, Uttar
Pradesh, India}

\date{today}%

\begin{abstract}

Several attempts have been made in the past decades to search for the true ground state of the dense matter at sufficiently large densities and low temperatures via compact astrophysical objects. Focusing on strange stars, we derive the hydrostatic equilibrium assuming a maximally symmetric phase of homogeneous superconducting quark matter called the \textit{color-flavor-locked} (CFL) phase in the background of energy-momentum squared gravity (EMSG). Theoretical and experimental investigations show that strange quark matter (SQM) in a CFL state can be the true ground state of hadronic matter at least for asymptotic densities, and even if the unequal quark masses. Motivated by these theoretical models, we explore the structure of stellar objects in recently proposed EMSG, which allows a correction term $T_{\mu \nu}T^{\mu \nu}$ in the action functional of the theory. Interestingly, EMSG may be effective to resolve the problems at high energy densities, e.g., relevant to
the early universe and dense compact astrophysical objects without invoking some new forms of fluid stress, such as bulk viscosity or scalar fields. Finally, we solve the complicated field equations numerically to obtain the mass-radius relations for strange stars in CFL equation of state.  

\end{abstract}


\maketitle

\section{Introduction}
Einstein’s General Relativity agrees with all tests in the solar system to a precision of $10^{-5}$ \cite{Will:2005va}. Cosmic acceleration led to two possibilities: exotic matter fields called dark energy, or a cosmological constant ($\Lambda$). However, some issues are still unresolved which keep open the way to frameworks which try to extend GR. These issues have led to another possibility by assuming that Einstein's GR has to be modified in some way. Hence, the search for modified gravity theories which may describe accelerating universe has become very popular due to their ability to provide an alternative framework to understand dark energy. Some of these modified theories are Lovelocks' theory of gravitation \cite{Lovelock,Lovelock:1972vz}, Einstein–Gauss–Bonnet theory \cite{Lanczos:1938sf}, $f(R)$ gravity \cite{Sotiriou:2008rp,DeFelice:2010aj}, etc. For a brief review of modified gravity theories, see Ref. \cite{DeFelice:2010aj}.
 
 In addition to the theories mentioned above, energy-mom-entum-squared gravity (EMSG) \cite{Arik:2013sti,mah1} has been proposed to encode the non-minimal coupling between geometry and matter. According to Ref. \cite{Arik:2013sti,mah1} , the Lagrangian contains an arbitrary functional of the Ricci scalar $R$ and the square of the energy-momentum tensor, i.e., $f(R, T^{\mu\nu}T_{\mu\nu})$ gravity. Interestingly, it has been found that the non-linear matter contributions in the field equations would affect the right-hand side of the EFEs without invoking some new forms of fluid stress, such as bulk viscosity or scalar fields. Concerning this approach, several interesting consequences have been reported, such as cosmological solutions
\cite{Akarsu:2017ohj,Akarsu:2018drb,Faria:2019ejh,Barbar:2019rfn}, dynamical system analysis \cite{Bahamonde:2019urw}, charged black hole  \cite{Chen}, wormhole solutions \cite{Moraes:2017dbs}, and so on. In addition to these studies, mass-radius relations of neutron stars have been studied for four different realistic EoS \cite{Akarsu:2018zxl}. In fact, authors have used recent observational measurements for the masses and radii of neutron stars to constrain the coupling constant $\alpha$. Further, in \cite{Maulana:2019sgd,mah2} polytropic EoS have been used to find mass-radius relation for neutron stars.

Neutron stars are dense, compact astrophysical objects with masses up to $2 M_{\odot}$ \cite{Demorest:2010bx,Antoniadis:2013pzd}. On the other hand, the radio pulsar PSR J1614-2230 \cite{Demorest:2010bx} around $1.97\pm 0.04$ $M_\odot$ mass has set rigid constraints on various matter EoS for neutron stars at high densities. Initially, it was assumed that neutron stars were composed of pure neutron matter described as a non-interacting relativistic Fermi gas. Current sensitivities put up constraints in the internal composition of neutron stars i.e. the composition and behaviour of equations of state (EoS) of the dense nuclear matter. In addition to this measurements on the radii of neutron stars provide additional constraints on the EoS \cite{Steiner:2012xt,Lattimer:2013hma}. However, in the aftermath of a core-collapse supernova explosion, several compact objects can sustain densities above a few times the nuclear saturation density in its interior. Thus, the composition and the properties of dense and strongly interacting matter is still an open question, and of the greatest importance for compact astrophysical objects. In spite of many efforts to explore the EoS, dense matter in the core of compact stars may consist of quark matter which is widely expected. On the other hand, several authors have considered an even more extreme possibility in the formation of a degenerate Fermi gas of quarks in which the quark Cooper pairs with very high binding energy condense near the Fermi
surface.  And their prediction is the Color-Flavor Locking (CFL) phase is the real ground state of Quantum Chromodynamics (QCD) at asymptotically large densities.\\
In this work, we focus on the CFL phase where all three flavors as well as three colors undergo pairing near the Fermi surface due to the attractive one-gluon exchange potential. In fact, the color neutrality constraint
is to be imposed in the CFL quark matter because a macroscopic chunk of quark matter must
be color singlet (see \cite{Alford2002,Steiner2002} for a review). 
 According to Ref. \cite{RAJAGOPAL2001,alford2007} quarks in the cores of neutron stars are likely to be in a paired phase.  Depending on the previous results one may consider that the CFL matter gives `absolute  stability' for sufficiently high densities \cite{Alford2004}. The CFL phase has several remarkable properties, 
such as CFL is more stable than SQM as long as $\mu \gtrsim$  $m_s^2/4 \Delta$, with $m_s$ being the strange quark mass and $\Delta$ the pairing gap \cite{Alford2001}; at asymptotically large densities the CFL phase is the energetically favored phase;  at extremely high densities, where the QCD gauge coupling is small, quark matter is always in the CFL phase with broken chiral symmetry and so on. As it was mentioned earlier in Ref. \cite{Flores:2017hpb,Flores:2010hpb,Banerjee:2020stc,Lugones:2002zd} that  CFL matter could be adequate candidate to explain stable neutron stars or strange stars.\\
From the above handful of literature it is clear that the structure of compact stars with CFL quark matter could represent a testbed for EMSG theory. The outline of the paper is the following: In Sec. \ref{sec2} we briefly introduce EMSG and its field equations. In Sec. \ref{sec3} we discuss the EoS for CFL strange matter. In Sec. \ref{sec4} we give a detailed analysis of the numerical methods employed to determine the mass-radius relations. Sec. \ref{sec5}, is devoted to reporting the general properties of the spheres in terms of the CFL strange quark matter. We finally draw our conclusion in Sec. \ref{sec6}.

\section{Field equations in energy-momentum squared gravity (EMSG)}\label{sec2}

The main feature of EMSG theory  is that the non-linear contributions of  EM tensor,
to encode the non-minimal matter-geometry coupling. The Lagrangian contains an arbitrary functional of the Ricci scalar and the square of the EM tensor, and the action for EMSG theory is \cite{Arik:2013sti,mah1}
\begin{equation}
S=\int \left({\frac{1}{8\pi}} ~\mathcal{R}+\alpha T_{\mu \nu}T^{\mu \nu}+\mathcal{L}_m \right) \sqrt{-g}~d^4x, \label{e1}
\end{equation}
where $\mathcal{R}$ is the Ricci scalar and $T_{\mu \nu}$ is the EM tensor with the coupling paramter $\alpha$. The 
 $\mathcal{L}_m$ denotes the matter Lagragian density.

The EMT, $T_{\mu \nu}$, is defined
via the matter Lagrangian density as follows
\begin{equation}\label{emt}
T_{\mu \nu} = -{2 \over \sqrt{-g}} {\delta (\sqrt{-g} \mathcal{L}_m) \over \delta g^{\mu \nu}}=\mathcal{L}_m~g_{\mu \nu}-2 {\partial \mathcal{L}_m \over \partial g^{\mu \nu}},
\end{equation}
which depends only on the metric tensor components,
and not on its derivatives.
If we vary the action \eqref{e1} with respect to $g^{\mu \nu}$, gives us the equation of motion for metric functions:
\begin{eqnarray}
\mathcal{R}_{\mu \nu}-{1 \over 2}\mathcal{R} g_{\mu \nu} = 8\pi T_{\mu \nu}+8\pi \alpha\left(g_{\mu \nu} T_{\beta \gamma}T^{\beta \gamma}-2\Theta_{\mu \nu} \right), \label{e3}
\end{eqnarray}
where,
\begin{eqnarray}
\Theta_{\mu \nu} &=& T^{\beta \gamma} {\delta T_{\beta \gamma} \over \delta g^{\mu \nu}}+T_{\beta \gamma} {\delta T^{\beta \gamma} \over \delta g^{\mu \nu}} \nonumber \\
&=& -2\mathcal{L}_m (T_{\mu \nu}-{1\over 2}g_{\mu \nu}T)-TT_{\mu \nu}+2T^\gamma_\mu T_{\nu \gamma}- \nonumber\\
&& 4T^{\beta \gamma} {\partial^2 \mathcal{L}_m \over \partial g^{\mu \nu} \partial g^{\beta \gamma}}
\end{eqnarray}
with $T=g^{\mu \nu}T_{\mu \nu}$, the trace of EMT.

Throughout this work we assume a perfect fluid EMT for the compact object. For that we assume $\mathcal{L}_m=P$ and using (\ref{emt}) the perfect 
fluid form is given by
\begin{eqnarray}
T_{\mu \nu} = (\rho+P)u_\mu u_\nu+Pg_{\mu \nu}, \label{e5}
\end{eqnarray}
where $\rho$ is the enrgy density, $P$ is the isotropic pressure with $u_\mu u^\mu=-1$ \& $\nabla_\nu u^\mu u_\mu=0$, respectively. The conservation equation can be found by covariant derivative of Eq. \eqref{e3}, which yield 
\begin{eqnarray}
\nabla^\mu T_{\mu \nu}=-\alpha g_{\mu \nu} \nabla^\mu (T_{\beta \gamma} T^{\beta \gamma})+2\alpha \nabla^\mu \Theta_{\mu \nu}. \label{e6}
\end{eqnarray}
Note that the standard conservation equation of
the energy-momentum tensor does not hold for this theory i.e., $\nabla^\mu T_{\mu \nu} $ is not identically zero.

After some algebra, one obtains the following field equations
\begin{eqnarray}
&& \hspace{-1cm} \mathcal{R}_{\mu \nu}-{1 \over 2}\mathcal{R} g_{\mu \nu} = 8\pi \rho \bigg[\left(1+{P \over \rho} \right)u_\mu u_\nu+{P \over \rho} g_{\mu \nu} \bigg]+8\pi \alpha \rho^2 \nonumber \\
&& \hspace{-0.5cm} \bigg[2\left(1+{4P \over \rho} + {3P^3 \over \rho^2} \right) u_\mu u_\nu + \left(1+{3P^2 \over \rho^2} \right)g_{\mu \nu} \bigg]. \label{e7}
\end{eqnarray}
The Eq. (\ref{e7}) can further reduce to coupled differential equations by consider a specific spacetime geometry. For the stellar configurations,  it is generally assume a spherically symmetric spacetime of the form
\begin{eqnarray}
\hspace{-0.5cm} ds^2 &=& - e^{2\nu} dt^2+e^{2\lambda} dr^2+r^2(d\theta^2+\sin^2 \theta ~d\phi^2), \label{e8}
\end{eqnarray}
with two independent functions $\nu(r)$ and $\lambda(r)$. Using the metric given in Eq. (\ref{e8}) in Eq. (\ref{e7}), we reach the following set of equations (see Ref. \cite{Akarsu:2018zxl})
\begin{eqnarray}
{e^{-2\lambda} \over r^2} \left(2r \lambda'-1 \right)+{1 \over r^2}= \rho_{\mbox{eff}}(r), \label{e9}\\
 {e^{-2\lambda} \over r^2} \left(2r \nu' +1 \right)-{1 \over r^2}  = P_{\mbox{eff}}(r), \label{e10}
\end{eqnarray}
where prime represent derivative with respect to $r$. Also, the effective
density and pressure $\rho_{\mbox{eff}}(r)$ and $P_{\mbox{eff}}(r)$ respectively, are given as
\begin{eqnarray}
\rho_{\mbox{eff}}(r)&=& 8\pi \rho +8\pi \alpha \rho^2\left(1+{8P \over \rho}+ {3P^2 \over \rho^2}\right), \nonumber \\
P_{\mbox{eff}}(r) &=&  8\pi P +8\pi \alpha \rho^2\left(1+{3P^2 \over \rho^2}\right). \nonumber 
\end{eqnarray}
To recast the Eq. (\ref{e10}) to a more familiar form we input  the gravitational mass function within the sphere of radius $r$, such that
\begin{eqnarray}
e^{-2\lambda} = 1-{2m(r) \over r}. \label{e11}
\end{eqnarray}
The other metric function, $\nu(r)$, is related to the pressure via
\begin{eqnarray}
\frac{d \nu}{dr} &=& -\left[\rho \left(1+{P \over \rho}\right) \left\{1+2\alpha \rho \left( 1+{3P \over \rho}\right) \right\} \right]^{-1} \nonumber \\
&& \Big[(1+6\alpha P) P'(r)+2\alpha \rho \rho'(r) \Big], \label{e12}
\end{eqnarray}
which is the radial component of the divergence of the field.
It is straightforward to use \eqref{e11} into \eqref{e9}, we have
\begin{eqnarray}
m'(r)=4 \pi  \rho  r^2 \left[1+\alpha  \rho  \left(\frac{3 P^2}{\rho ^2}+\frac{8 P}{\rho }+1\right)\right]. \label{e13}
\end{eqnarray}
Finally, using the expressions \eqref{e9}-\eqref{e12}, the modified TOV equations take the following convenient form \cite{Akarsu:2017ohj}
\begin{eqnarray}
P'(r) &=& -\frac{m \rho }{r^2} \left(1+\frac{P}{\rho }\right) \left( 1-\frac{2 m}{r}\right)^{-1} \bigg[1+\frac{4 \pi  P r^3}{m}+ \alpha \nonumber \\
&& \frac{4 \text{$\pi $r}^3 \rho ^2}{m} \left(\frac{3 P^2}{\rho ^2}+1\right) \bigg] \bigg[1+2 \alpha  \rho  \left(1+\frac{3 P}{\rho }\right) \bigg] \nonumber \\
&& \bigg[1+2 \alpha  \rho  \left({d\rho \over dP}+\frac{3 P}{\rho }\right) \bigg]^{-1}. \label{e14}
\end{eqnarray}
The system of Eqs. \eqref{e9}-\eqref{e14} are
not enough to solve for the four variables, since there are one degrees of freedom. To complete this set of equations, we need now to specify the EoS relating the pressure and energy density of the  fluid.

\section{Color-flavor locked  equations of state} \label{sec3}

Here, we outline the equation of state (EoS) for CFL quark matter that can be obtained in the framework of the MIT bag model. In the CFL phase, the thermodynamic potential for  electric and color charge neutral CFL quark matter is given by \cite{alf01} 
\begin{eqnarray}
 \Omega_{CFL}&=& -{3\Delta^2 \mu^2 \over \pi^2}+{6 \over \pi^2} \int_0^{\gamma_F} p^2 (p-\mu)~dp   \nonumber \\
&+&{3 \over \pi^2} \int_0^{\gamma_F} p^2 \left(\sqrt{p^2+m_s^2}-\mu \right)~dp+B, ~~~~~~~~\label{cfl}
\end{eqnarray}
to the order of $\Delta^2$, where $\mu$ is the quark chemical potential and $\Delta$ denotes the color superconducting gap parameter of CFL phase of quark matter. The first term is the contribution of the CFL condensate to $\Omega_{CFL}$, while the 
second and third terms of (\ref{cfl}) give the thermodynamic potential of (fictional) unpaired quark matter in which
all quarks that are going to pair have a common Fermi momentum $\gamma_F$ which minimizes the thermodynamic potential of the fictional unpaired quark matter \cite{Alford:2002rj}. The final term is the bag constant.

The common Fermi momentum is given by 
\begin{eqnarray}
\gamma_F = \bigg[\bigg( 2\mu - \sqrt{\mu^2+{m_s^2-m_u^2 \over 3}}\bigg)^2 -m_u^2 \bigg]^{1/2}
\end{eqnarray}
where $\mu=(\mu_s+\mu_u+\mu_d)/3$ i.e. the average quark chemical potential, $m_s ~\&~ ~m_u$ are  strange and up quark masses respectively.  For massless up and down quarks we get
\begin{eqnarray}
\gamma_F &=& 2\mu - \sqrt{\mu^2+{m_s^2 \over 3}} \sim \mu-{m_s^2 \over 6\mu}.  
\end{eqnarray}

By following the pairing ansatz in the CFL phase \cite{Steiner2002} 
\begin{eqnarray}
n_u=n_r, ~~~~~ n_d = n_g, ~~~~~\text{and}~~~~~ n_s = n_b
\end{eqnarray}
where $n_r$, $n_g$, $n_b$ and $n_u$, $n_d$, $n_s$ are color and flavor number densities respectively.  In the discussion that follows, color neutrality automatically enforces electric charge neutrality in the CFL phase, and the quark number densities are $n_u = n_d = n_s = {\gamma_F^3+2\Delta^2 \mu \over \pi^2}$. It is to be noted that the color neutral CFL quark matter is electric charge neutral, the corresponding electric charge
chemical potential is $\mu_e= 0$. In our discussion we consider the values of the CFL gap parameter in the range
of $\Delta \sim 50-100 ~MeV$ (see Ref. \cite{Alford1998,Rapp1998}. As the necessary condition for
MIT based EoS the bag constant $B$ to be always greater than 57 $MeV/fm^3$ \cite{far84}. In fact, the free energy contributed from CFL pairing is more than the free energy consumes to maintain equal number of quark densities \cite{alf01}. Thus,  CFL paired quarks are more stable than unpaired.\\

Since it is always difficult to obtain an exact expression for an EoS when $m_s \neq 0$.  However,  a simple EoS similar to the MIT-bag model can be obtained for $m_s=0$, with an extra term from CFL contribution as $\rho = 3p+4B-6\Delta^2 \mu^2 / \pi^2$. Considering the series upto the order $\Delta^2$ and $m_s^2$, the expression for pressure and energy density in the CFL phase can be obtained as \cite{lugo02} 
\begin{eqnarray}
P &=& {3\mu^4 \over 4\pi^2}+ {9\beta \mu^2 \over 2\pi^2}-B ~,~\mbox{and}~\rho = {9\mu^4 \over 4\pi^2}+ {9\beta \mu^2 \over 2\pi^2}+B, \notag \\
\end{eqnarray}
where $\beta = -m_s^2/6+2\Delta^2/3$.  Finally, an explicit function of the energy density $\rho$ in the form
\begin{eqnarray}
\rho = 3P+4B-{9\beta \over \pi^2} \left\{\left[{4\pi^2 (B+P) \over 3}+9\beta^2 \right]^{1/2}-3\beta \right\}. \nonumber \\\label{e18} 
\end{eqnarray}

\section{Numerical Approach}\label{sec4}
Since the field equations \eqref{e9} and \eqref{e10} are highly non-linear, so we adopt numerical integration in Mathematica. For solving the field equations we will be directly use the TOV-equation \eqref{e14} along with equation of mass function \eqref{e13}. Since the two equations include three unknown quantities i.e. $\rho(r), ~P(r)$ and $m(r)$, we need an additional information. Therefore, we will consider an EoS for color-flavor locked (CFL) quark matter given in \eqref{e18} that generalizes the MIT bag model. As a first step we convert the units in EoS, TOV and mass function equations so that the mass will be measured in solar mass, radius in km, pressure, density  \&  bag constant $B$ in $MeV/fm^3$, strange quark mass $m_s$ \& color-superconducting gap $\Delta$ in $MeV$.  \\

\begin{figure}
\centering
\includegraphics[width=1\linewidth]{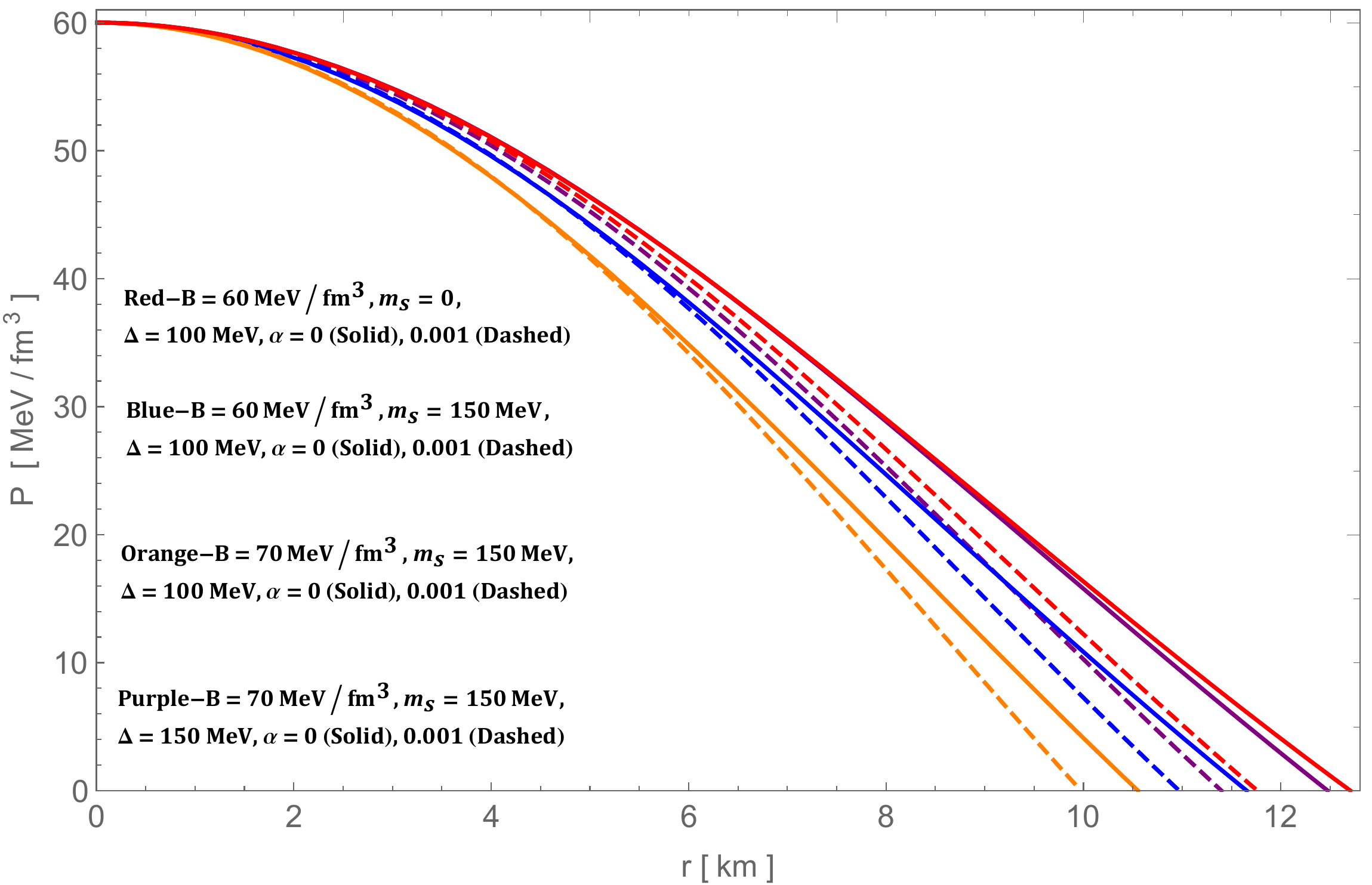}
\caption{Variation of pressure with radius for different $(B,m_s,\Delta,\alpha)$ and $P(r_0)=60~MeV/fm^3$.}
\label{f1}
\end{figure}

\begin{figure}
\centering
\includegraphics[width=1\linewidth]{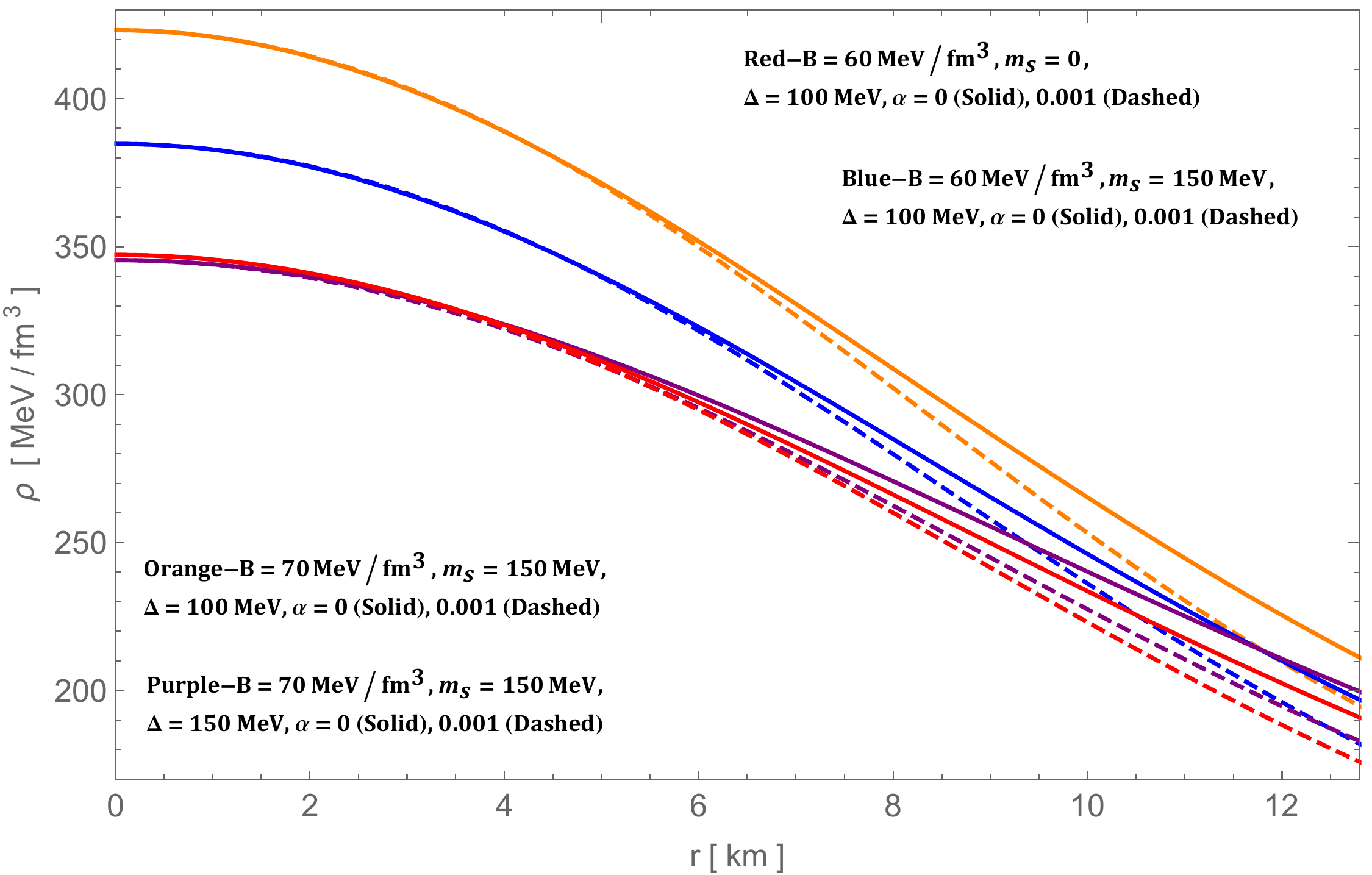}
\caption{Variation of energy density with radius for different $(B,m_s,\Delta,\alpha)$ and $P(r_0)=60~MeV/fm^3$.}
\label{f2}
\end{figure}

Next we use ``\texttt{NDSolve}'' package in Mathematica defining a coupled differential equations \eqref{e13} and \eqref{e14} with initial ($\mathcal{I}$) and boundary ($\mathcal{B}$) conditions given below:
\begin{eqnarray}
\mathcal{I} &:& P(r_0)=p_0, ~~~r_0=10^{-10}\,km\nonumber \\
&& m(r_0) = {4\pi r_0^3 \over 3} \rho_0 \left[1+\alpha  \rho_0 \left(\frac{3 p_0^2}{\rho_0^2}+\frac{8 p_0}{\rho_0}+1\right) \right]~~~~~ \\
\mathcal{B} &:& P(R) = 0, ~~~~~~ m(R)=M,
\end{eqnarray}
and solve for pressure and mass functions. Here $P(r_0)=p_0$, $\rho(r_0)=\rho_0$, $R$ is the radius of the star and $M$ is the total gravitational mass. To proceed we start by supplying the values of the constant parameters i.e. $(B,m_s,\Delta,\alpha)$. {Solving the TOV equation by choosing a central pressure e.g. $P(r_0)=60~MeV/fm^3$ until the pressure vanishes i.e. $P(R)=0$, which defines the surface of the star generating the Figs. \ref{f1}-\ref{f5}. The Figs. \ref{f6}-\ref{f9} were generated by integrating the TOV-equation for different values of $P(r_0)$}. In the following, we will focus on the four cases: (i) $[60~MeV/fm^3,~0,~100MeV]$, (ii) $[60~MeV/fm^3,~150MeV$ $,~100MeV]$, (iii) $[70~MeV/fm^3,$ $150MeV,~100MeV]$, and (iv) $[70~MeV/fm^3,~150MeV,$ $~150MeV]$, respectively.  Further, we analyse the resulting mass and radius for the central pressure allowed by the EoS under consideration, and each cases have been examined carefully which are valid from a physical point of view.

\section{Physical acceptability of the model} \label{sec5}
To check the numerical solution for its acceptability through physical constraints, we need to analyze thoroughly and how it behaves when changing the constant parameters.

\subsection{Non-singular central density and pressure}
In order to check the EoS behavior and viability of a compact star model,  the central values of density and pressure must be finite. Since the central pressure is one of the initial input parameters for the numerical integration, it is clearly finite. Figure \ref{f1} shows the variation of pressure for different physical inputs in the interior of compact objects. Then solving the pressure from TOV-equation, the density can be calculated using EoS. From the Fig. \ref{f2}, we can also see that the central densities for 4 cases are finite and thereby both the central density and pressure are non-singular.\\

It can also be seen that as the strange quark mass increases  ($0 \rightarrow 150 ~MeV$), the radius of the star decreases while the central density increases, see  Red \& Blue curves in Figs. \ref{f1} and \ref{f2}, respectively. Further, when bag constant increases  ($60 \rightarrow 70 ~MeV/fm^3$), the radius of the star reduces significantly while the density increases very much (Figs. \ref{f1}, \ref{f2}, Blue \& Orange). Again, if the color superconducting gap increases ($100 \rightarrow 150 ~MeV$), the radius of the star increases as compared to the case III but still lesser than case I while almost equivalent with case II. However, the density is lower than case III but slightly higher than case II. Finally, the comparison of GR ($\alpha=0$) and EMSG ($\alpha=0.001$) is that GR has larger surface boundary than EMSG counterpart while the density is the same for both gravity till about 6 km than start lesser value in EMSG than GR i.e. EMSG has lower surface density than GR counterpart.




\begin{figure}
\centering
\includegraphics[width=1\linewidth]{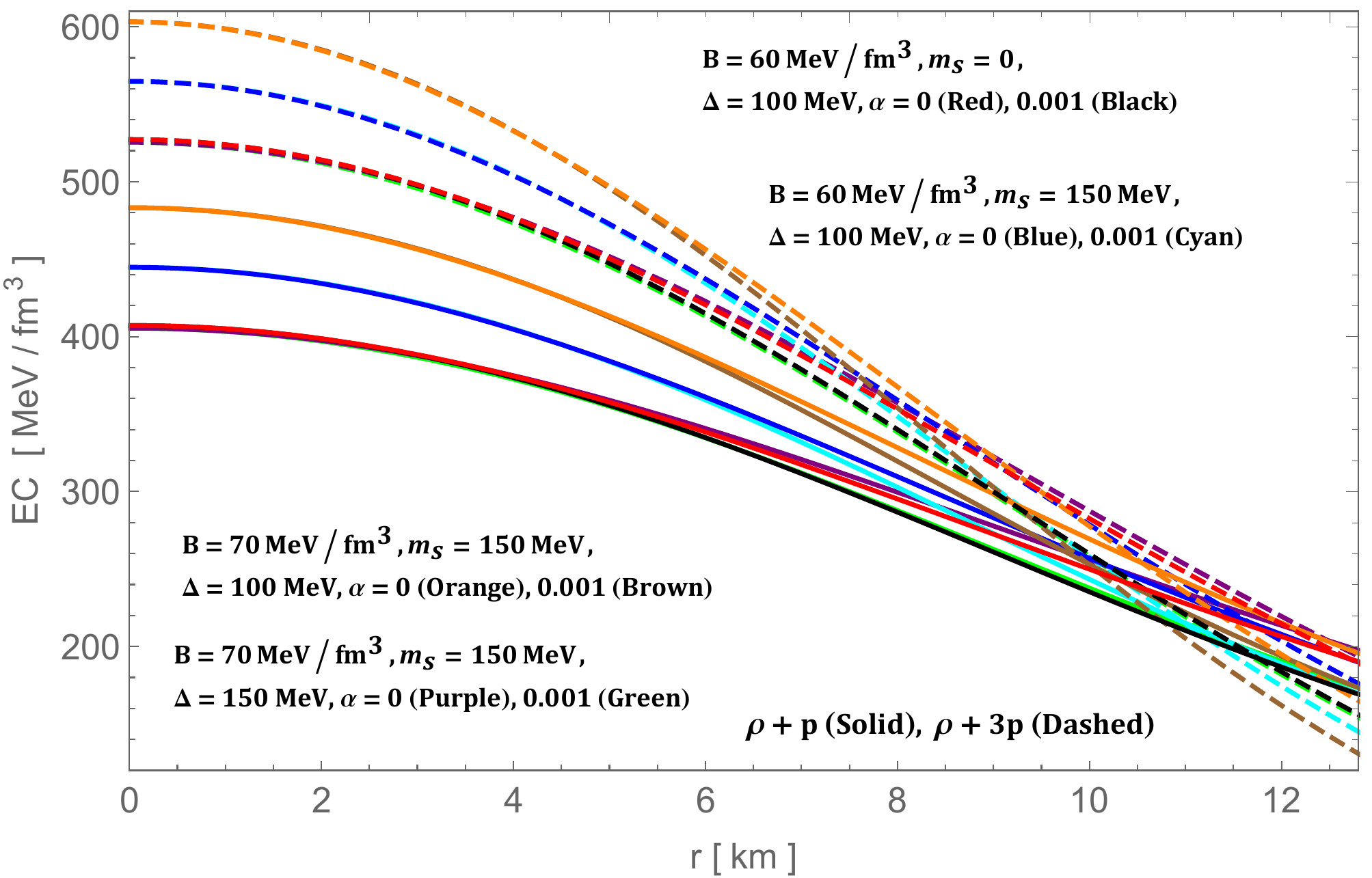}
\caption{Variation of energy conditions with radius for different $(B,m_s,\Delta,\alpha)$ and $P(r_0)=60~MeV/fm^3$.}
\label{f5}
\end{figure}

\subsection{Energy conditions}
For any physically plausible fluid, physical constraint demands strict energy conditions namely strong, weak, null and dominant conditions or mathematically
\begin{eqnarray}
\text{SEC} &:& \left(T_{\mu \nu}-{1 \over 2}T g_{\mu \nu} \right)u^\mu u^\nu \ge 0 ~~~\text{or}~~~\rho+P\ge 0, \nonumber \\
&& \hspace{1cm} \rho+3P \ge 0,\\
\text{WEC} &:& T_{\mu \nu} u^\mu u^\nu \ge 0, ~~\text{or}~~ \rho \ge 0,~~\rho+P \ge 0,\\
\text{NEC} &:& T_{\mu \nu} k^\mu k^\nu \ge 0 ~~\text{or}~~ \rho+P \ge 0,\\
\text{DEC} &:& T_{\mu \nu} v^\mu v^\nu \ge 0 ~~\text{or}~~ \rho \ge |P|,
\end{eqnarray}
where $u^\mu$ is time-like vector, $k^\mu$ is the null-vector and $v^\mu$ is any future directed causal vector. All these energy conditions is fulfilled by the solution from Figs. \ref{f1}, \ref{f2} and \ref{f5}.

\begin{figure}
\centering
\includegraphics[width=1\linewidth]{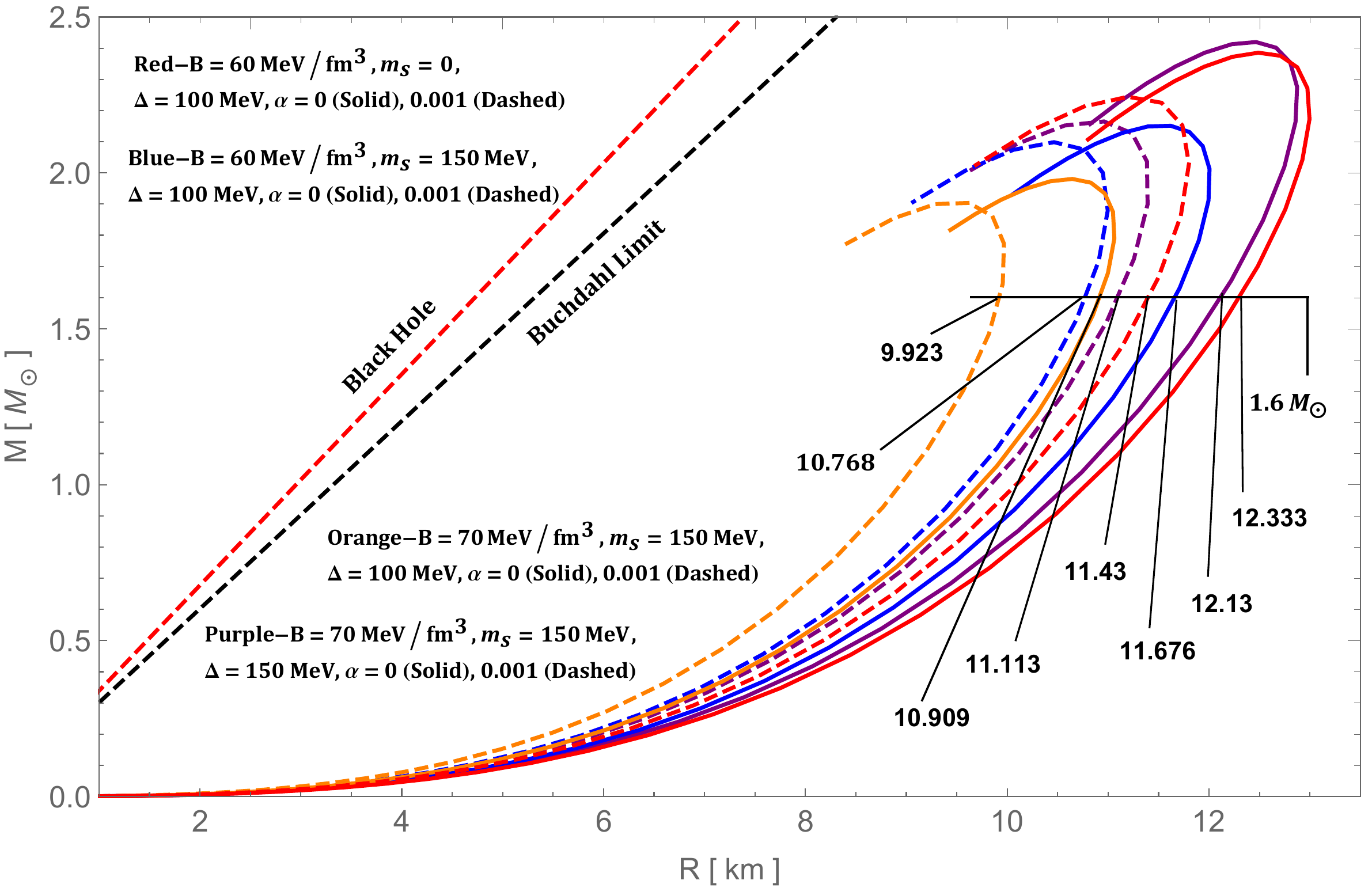}
\caption{$M-R$ for different $(B,m_s,\Delta,\alpha)$ {by varying $P(r_0)$}.}
\label{f6}
\end{figure}

\begin{figure}
\centering
\includegraphics[width=1\linewidth]{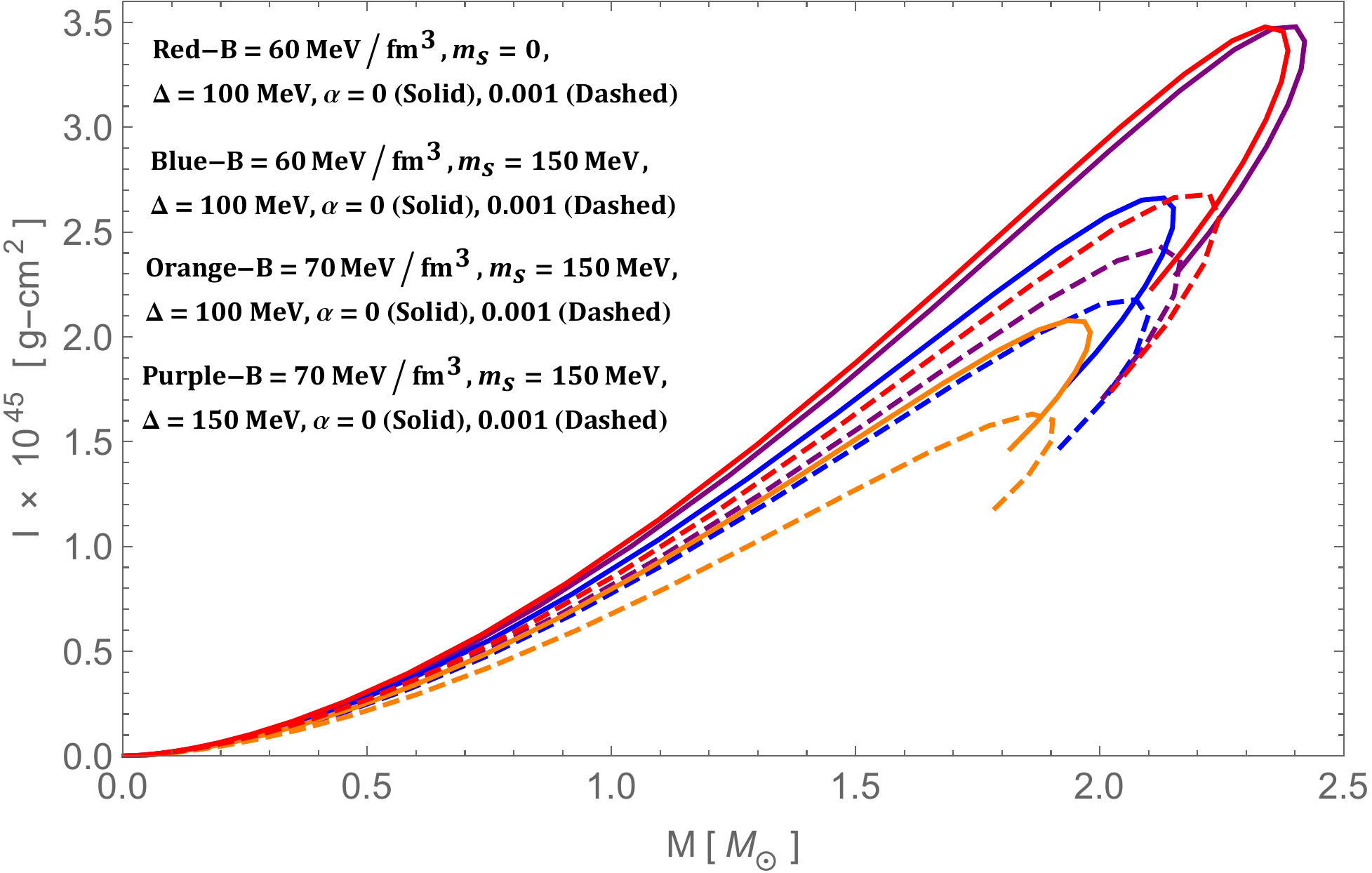}
\caption{$M-I$ for different $(B,m_s,\Delta,\alpha)$ {by varying $P(r_0)$}.}
\label{f7}
\end{figure}

\subsection{$M-R$ and $M-I$ curves}

The $M-R$ curve was directly generated from the TOV-equation via the numerical solution. This only means that EMSG, increasing $B$ and $m_s$ soften the EoS while increasing $\Delta$ stiffens it. {This graph was generated by varying the central pressure $P(r_0)$ and measuring the mass with the corresponding radius at the boundary $P(R)=0$.} For the cases, introducing EMSG soften the EoS thereby reducing the maximum mass (Solid \& Dashed lines).  From Fig. \ref{f6}, one can observed that increasing strange quark mass reduces the $M_{max}$ (Red \& Blue) and further $M_{max}$ reduces when increasing bag constant (Blue \& Orange) however, for the same bag constant when color superconducting gap increases the $M_{max}$ significantly increases. In the $M-R$ curve Fig. \ref{f6}, we have also included the Buchdahl condition where the collapse of the star may proceed beyond it and ultimate form a black hole once the $r=2m$ is fulfilled.

The $M-I$ curve is obtain by using the approximate relationship which is defined as \cite{bejg}
\begin{eqnarray}
I={2 \over 5} \left(1+{M \over R}\, {km \over M_\odot} \right). \label{e27}
\end{eqnarray}
Equation \eqref{e27} has the accuracy of 5\% and less and on using this equation we have generated the $M-I$ curve (Fig. \ref{f7}).

\begin{figure}
\centering
\includegraphics[width=1\linewidth]{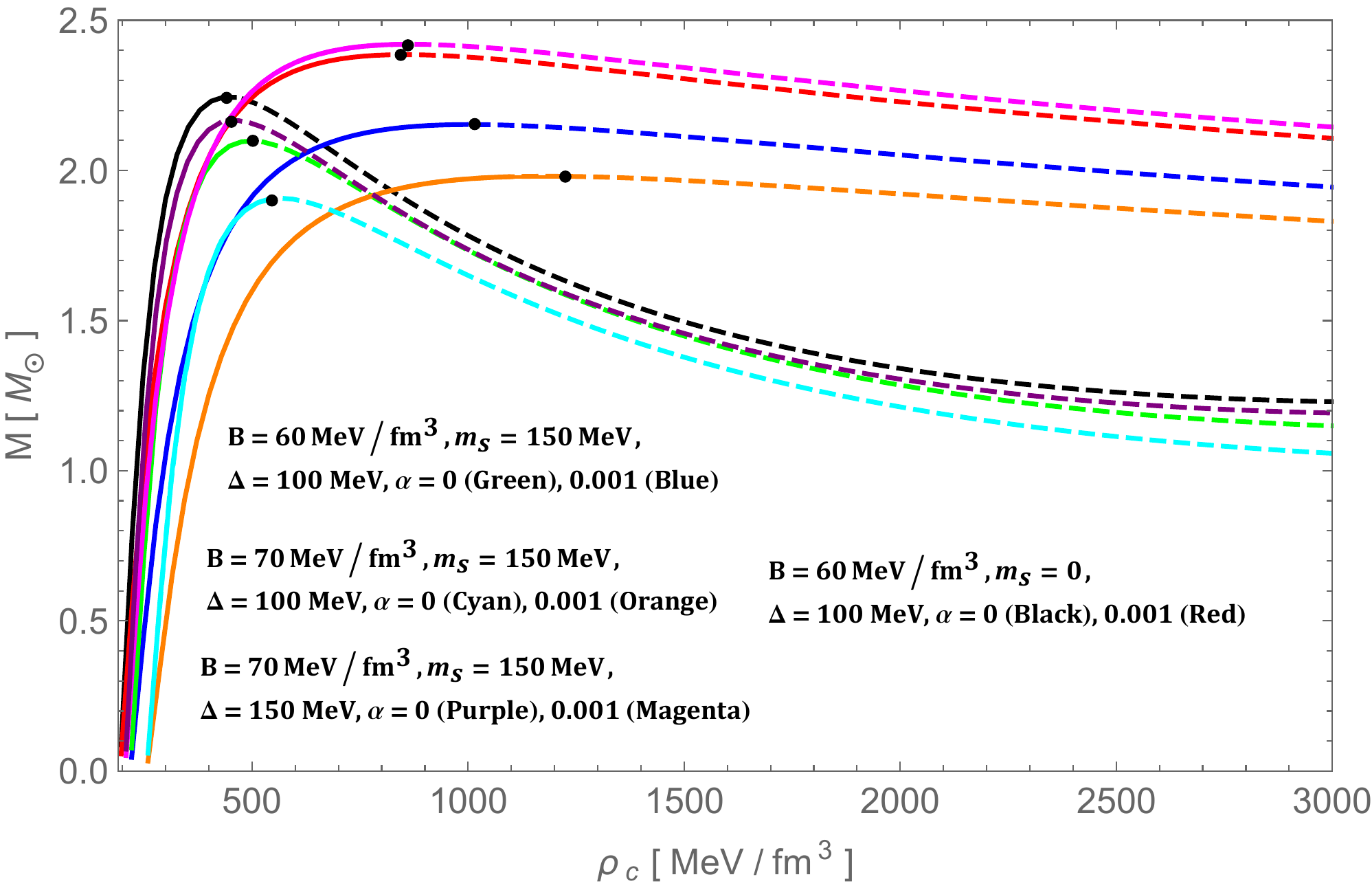}
\caption{$M-\rho_c$ for different $(B,m_s,\Delta,\alpha)$ {for different values of $P(r_0)$}.}
\label{f8}
\end{figure}

\begin{figure}
\centering
\includegraphics[width=1\linewidth]{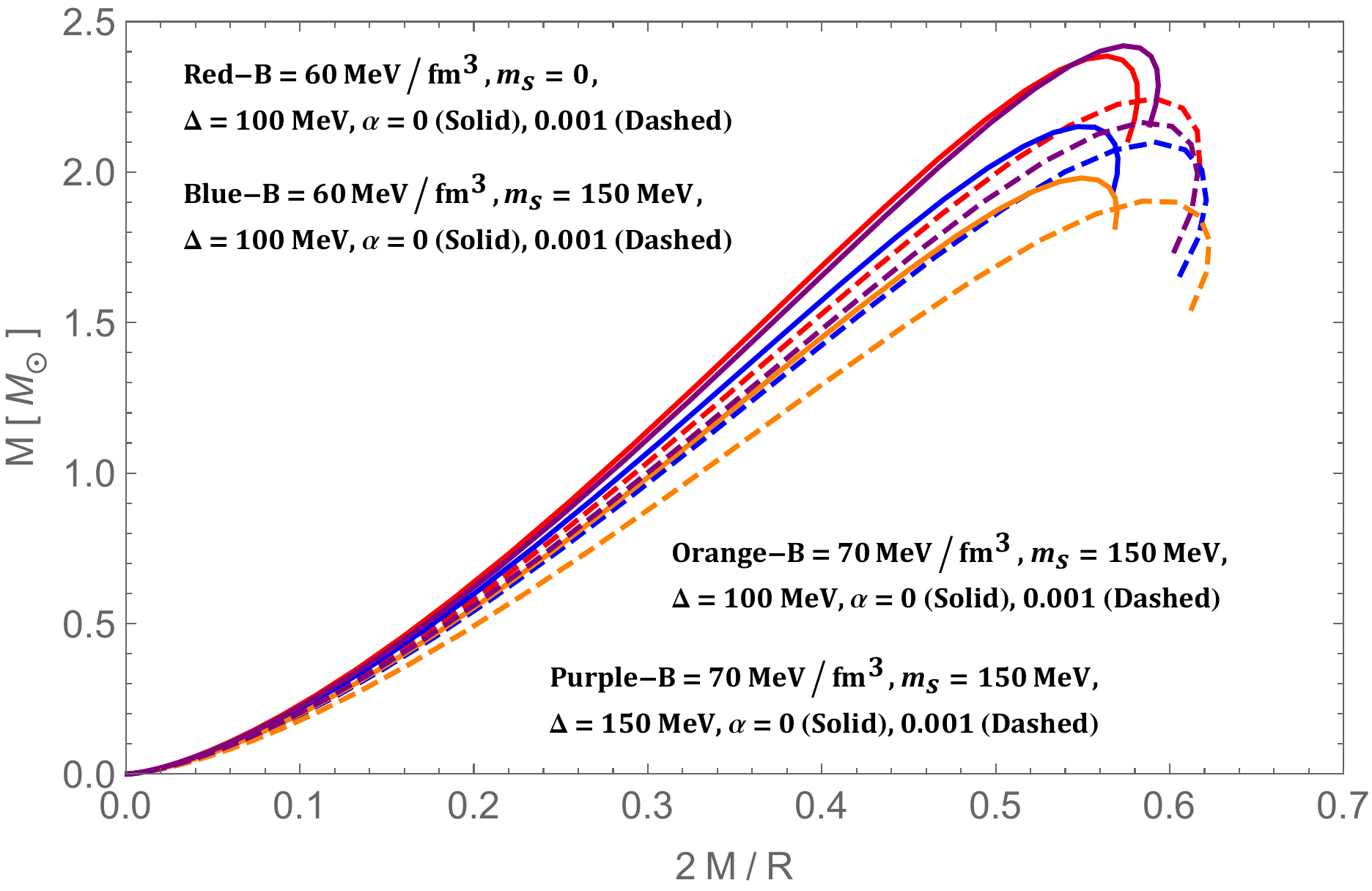}
\caption{$M-2M/R$ for different $(B,m_s,\Delta,\alpha)$ {for different values of $P(r_0)$}.}
\label{f9}
\end{figure}

\section{Static stability criterion}

{For completeness, we would also like to check the stability of compact stars constructed in EMSG and explore the physical properties in the interior of the fluid sphere. In the following, we discuss the mass with central energy density, compactness, and the stability of stars.}

\subsection{Mass with the central energy density}
Once the TOV stationary configuration has been determined, one can probe its stability towards collapse. A necessary (but not sufficient) condition for stability of a compact star is that the total mass is an increasing function of the central density, $d M/d \rho_c >0$ \citep{Zeldovich:1971,Harrison:1965ag};  this is the {\it {static stability criterion}}.  Therefore, it is interesting to ask what would happen to our constraints 
on $M_{\text{max}}$ if we measured a quark star with  $\sim 2M_{\odot}$.  
Our mass-central density diagram is reported in Fig. \ref{f8} for the stellar configuration obtained by solving the TOV equations and changing the value of central density. Here, we want to stress that while performing the calculation we have chosen {different values of central pressure while the corresponding central density is found to be in the range $340< \rho_c< 450 ~MeV/fm^3$ for different values of  constant parameters, see Fig. \ref{f2}}. From Fig. \ref{f8}, one can immediately see that the mass initially increases with central density until it attains a maximum value. Then the mass decreases with the increasing central density. The most interesting part is that in EMSG (i.e. $\alpha \neq 0$ ), all curves having  positive slops i.e.  $d M/d  \rho_c >0$ in the regime, $340< \rho_c< 450 ~MeV/fm^3$. But outside this range  most of them correspond to the unstable configuration.  Therefore, this solution presents a very dense stable star upto certain range depending on the values of central density and pressure. Obviously, this is a necessary condition for a stable equilibrium configuration. Interestingly, the solid curves corresponding to GR (i.e. $\alpha = 0$ ) are quite satisfactory and  consequently leads to a stable configuration. 
{Of course, a careful and detailed analysis of stability conditions in EMSG, would help to clarify this issue. We leave this point as a subject of study for future works.}

\subsection{Compactness}

{It is useful to explore extremities of compactness for compact star structure.  Thus, we introduce the dimensionless compactness parameter $r_{g}/R= 2M/R$, where $r_{g}$ is the Schwarzschild radius. This situation is illustrated in Fig. \ref{f9} for different  values of parameters $(B,m_s,\Delta,\alpha)$.  Regarding the stars with the same $M$ and coupling constant $\alpha$, it is noticed that the compactness increases when the bag constant increases. Here we define the Schwarzschild radius $r_{g}=2M$.   The detailed behavior of compactness is shown in  Fig. \ref{f9} which  lies in the range  $0.54<r_{g}/R< 0.62$ for QSs corresponding their respective EoS.}

\section{Results and discussions}\label{sec6}

In the present paper we have tested energy-momentum squared gravity (EMSG) model in the strong gravity field regime  using quark stars. We have successfully presented a quark star model that quite accurately meets the observed  constraints on the mass-radius relation and corresponding maximal mass. Moreover, the recent observations of gravitational waves (GW) from binary NS mergers GW 170817 \cite{TheLIGOScientific:2017qsa} by the LIGO-Virgo Collaboration (LVC) has opened up a new avenue to constrain the EOS at high densities. Here, we assumed the equation of state for CFL quark matter that can be obtained in the framework of the MIT bag model. Using this EoS we solved the system of equations and derived the mass-central density  and mass-radius profile for quark stars numerically.

Our main motivation is to discuss the features of the EMSG model and the affects of ($B,$ $m_s,~\Delta,~\alpha$) on the physical properties of the CFL star in these theories. Then, we discuss the physical implications of such compact object by analyzing the trends of pressure, energy density and  mass-radius relation (Figs. \ref{f1},\ref{f2},\ref{f6}), respectively. The decreasing nature of the pressure and energy density are one of the important requirements for realistic compact star models.  Moreover, the fulfilment of energy conditions makes CFL star composed of non-exotic matters.

We further extend our article by discussing the hydrostatic equilibrium equations in spherical symmetry from the field equations within the EMSG context. We discussed the static stability criterion which is required to satisfy by all stellar configurations (see Fig. \ref{f8}). We also examined the mass-radius relation depends on the choice of the value for the coupling constant $\alpha$. Interestingly, the CFL quark matters can support maximum mass when the strange quark mass $m_s$  is negligible in GR case. When the EMS coupling is introduced the CFL matter get softer thereby reducing the maximum mass. Further, if $m_s$ increases the stiffness of CFL matter gets more softer, again the $M_{max}$ is reduced further. However, when the bag constant increases the stiffness of the CLF matters significantly reduces. Although, with the increase in color superconducting gap $\Delta$ the stiffness also increases significantly. These effects due to changes in $(B,m_s,\Delta)$ are similar to Lugones \& Horvath predictions \cite{lug} however, the coupling constant $\alpha$ modification in EMSG action makes the physical properties changes drastically. In all the cases, the radius for a particular mass is always lesser in EMSG than GR thereby more compactness factor. Due to more compactness factor the surface redshift $z_s=(1-2M/R)^{-1/2}-1$ will be more in EMSG than GR. Also in Fig. \ref{f8}, it is clear that the mass saturate faster with smaller range of central density in EMSG (Dahsed Lines) and saturate at larger $\rho_c$ range in GR (Solid Lines). These means that the stellar system will be more stable in GR than EMSG when changing the central density due to radial perturbations.

 On the other hand, we have seen in Ref. \cite{baus} that using the data from GW 170817 the radius of a neutron star $1.6M_\odot$ must be above $10.68^{+0.15}_{-0.03}$ km.
Followed by this, we see that the radius corresponding to $1.6M_\odot$ in $M-R$ curve ( Fig. \ref{f6}) is in agreement with this constraints except for a case with $B=70MeV/fm^3,~m_s=150MeV,~\Delta=100MeV,~\alpha=0.001,$ whose radius is $9.923$ km. This means that our model is physically realistic and might have astrophysical applications in future.

\subsection*{Acknowledgments}
S. K. Maurya is thankful for continuous support and encouragement from the administration of University of Nizwa.

\end{document}